\RequirePackage{xspace}

\def\ifundefined#1{\expandafter\ifx\csname#1\endcsname\relax}

\def\la{\mathrel{\hbox{\rlap{\hbox{\lower4pt\hbox{$\sim$}}}\hbox{$<$}}}}
\def\ga{\mathrel{\hbox{\rlap{\hbox{\lower4pt\hbox{$\sim$}}}\hbox{$>$}}}}

\newcommand{\be}{\begin{equation}}
\newcommand{\ee}{\end{equation}}
\newcommand{\bea}{\begin{eqnarray}}
\newcommand{\eea}{\end{eqnarray}}
\ifundefined{ensuremath}\def\ensuremath#1{\relax\ifmmode{#1}}
\else${#1}$\fi\else\relax\fi
\ifundefined{nuc}\def\nuc#1#2{\relax\ifmmode{}^{#1}{\protect\textrm{#2}}
\else${}^{#1}$#2\fi}\else\relax\fi

\newcommand{\kmps}{\ensuremath{\mathrm{km}~\mathrm{s}^{-1}}\xspace}

\newcommand{\msol}{\ensuremath{{\mathrm{M}_\odot}}\xspace}

\newcommand{\phx}{\texttt{PHOENIX}\xspace}

\newcommand{\RSiSu}{\ensuremath{\Re_{SiS}}} 
\def\lstar{\ifmmode{\Lambda^*}\else\hbox{$\Lambda^*$}\fi} 
\def\Lstar{\ifmmode{\Lambda^*}\else\hbox{$\Lambda^*$}\fi} 
\def\Rop{\ifmmode{[R_{ij}]}\else\hbox{$[R_{ij}]$}\fi}

\def\Rji{\ifmmode{[R_{ji}]}\else\hbox{$[R_{ji}]$}\fi}
\def\Rstar{\ifmmode{[R_{ij}^*]}\else\hbox{$[R_{ij}^*]$}\fi}

\def\Rjistar{\ifmmode{[R_{ji}^*]}\else\hbox{$[R_{ji}^*]$}\fi}
\def\DRji{\ifmmode{[\Delta R_{ji}]}\else\hbox{$[\Delta R_{ji}]$}\fi}
\def\DRij{\ifmmode{[\Delta R_{ij}]}\else\hbox{$[\Delta R_{ij}]$}\fi}

\def\Jb{{\bar J}}

\def\e{\epsilon}
\def\Jb{{ J}}
\def\Jnew{{ J_{\rm new}}}
\def\Jold{{ J_{\rm old}}}
\def\Jfs{{ J_{\rm fs}}}
\def\Snew{{S_{\rm new}}}
\def\Sold{{S_{\rm old}}}
\def\lstar{\ifmmode{\Lambda^*}\else\hbox{$\Lambda^*$}\fi} 
\def\Lstar{\ifmmode{\Lambda^*}\else\hbox{$\Lambda^*$}\fi}

\documentclass[
  ,draft              ]
  {aipproc}
  
\layoutstyle{6x9}

\graphicspath{{./}{./epsfigs/}}

\begin{document}

\title{3D Radiative Transfer with PHOENIX}

                

\author{E. Baron}{
  address={Homer L.~Dodge Department of Physics and Astronomy,
             University of Oklahoma, 440 W Brooks, Rm 100, Norman, OK
             73019-2061 USA}
  ,altaddress={        Computational Research Division, Lawrence Berkeley
        National Laboratory, MS 50F-1650, 1 Cyclotron Rd, Berkeley, CA
        94720 USA
}
} 
\author{Bin Chen}{
  address={Homer L.~Dodge Department of Physics and Astronomy,
             University of Oklahoma, 440 W Brooks, Rm 100, Norman, OK
             73019-2061 USA}
}

\author{Peter H.~Hauschildt}{
  address={Hamburger Sternwarte, Gojenbergsweg 112, 21029 Hamburg, Germany}
}

\begin{abstract}
Using the methods of general relativity Lindquist derived the
radiative transfer equation that is correct to all orders in
$v/c$. Mihalas developed a method of solution for the important case
of monotonic velocity fields with spherically symmetry. We 
have developed the generalized atmosphere code \phx, which in 1-D has
used the framework of Mihalas to solve the radiative transfer equation
(RTE) in 1-D moving flows. We describe our
recent work 
including 3-D radiation transfer in \phx and 
particularly including moving flows exactly using a
novel affine method. We briefly discuss quantitative spectroscopy in
supernovae.  
\end{abstract}

\maketitle

\section{Introduction}

In the  1970s Dimitri Mihalas and collaborators developed the
equations of radiative transfer and numerical techniques for their
solution \citep{mih80,mk78,mkh76a,mkh76b,mkh76c,mkh75} in
relativistically expanding atmospheres that are required for the
quantitative spectroscopic modeling of supernovae (SNe). Over the last
twenty years our group has developed \phx to include most of the
physics that is needed to calculate supernova atmospheres. While many
of the techniques in \phx
 are modern \citep{phhcas93,phhre92,phhs392,hbjcam99,hbmathgesel04},
the underlying solution of the equations follows the method developed
by \citet{mih80}. While this method works well for spherically
symmetric flows, it is difficult to extend it to the case of fully 3-D
atmospheres even with the restriction of monotonic velocity  fields (see the
contribution by J.~I.~Castor, this volume). Here we describe a method
of using an affine parameter in order to calculate the solution along
geodesics (straight lines in flat spacetime). This method is
straightforward, exact to all orders in $v/c$, and can be generalized
to arbitrary flows and include the effects of curved spacetime.

\section{Motivation}

Why study supernovae? Supernovae are common events that 
occur about once a second out in the volume out to redshift $z \approx 1$. 
Supernovae play a major role in galactic nucleosynthesis. 
Supernovae inject energy into ISM, and  trigger star formation.
And finally, SNe Ia make good standardizable candles. It is important
to emphasize while the standard candle relation is purely empirical
\citep{philm15,philetal99} it is understood theoretically, at least
qualitatively as being due to higher temperatures due to nickel mass
variations \citep{kmh93,hofkhoklc96,nugseq95} and opacity variations
\citep{kmh93,KW07}.

Observationally, supernovae are classified by their spectra and there
are a large number of classifications which are described in the review
article by \citet{filarev97}. Theoretically, there are two supernova
mechanisms, core collapse and thermonuclear. Core collapse occurs at
the end of the life of a massive ($M> 8$~\msol) star, when the iron
core collapses to nuclear matter density and then bounces. Exactly how
the shock gets out of the iron core is a subject of current
research. The wide variety in observed spectra is thought to be due to
variations in the progenitor (does it have an intact hydrogen envelope
or has it lost some or all of its envelope either in a wind or via
interaction with a companion). Thermonuclear supernova refers to the
thermonuclear burning 
of a C+O white dwarf that accretes mass from a companion until it
reaches the Chandrasekhar mass, these objects form the class SNe
Ia. Since most, in not   all, SNe Ia explode at the same mass this naturally
accounts for the relatively small amount of variation in their
intrinsic brightness at maximum light.
Since SNe Ia are quite bright they  can be found far away. The
small variation in brightness combined with their high intrinsic
brightness makes them excellent cosmological probes. 

The small intrinsic variation of SNe Ia at maximum light
can be corrected for using  the light curve shape method. This follows
from the realization
that the luminosity at peak is correlated with the rate at which the light
curve declines from maximum light \citep{philm15}. Refining the
original suggestion of Phillips has led to improved light curve shape
parametrizations \cite{philetal99,goldhetal01,JRK07}. Using SNe Ia as
correctable candles, two groups discovered the dark energy
\cite{riess_scoop98,perletal99}. As noted above the light curve shape
relations are purely empirical and determined using nearby
supernovae. Thus, if the distant sample is significantly different
from the nearby sample this could result in large systematic errors in
the results for the nature of the dark energy (the existence of the
dark energy seems to be on pretty solid ground). 
One way to search for differences is to compare spectra of the nearby
sample to those in the distant sample. \citet[][see their Figure
11]{riess_scoop98}  found that the nearby sample was pretty similar to
the distant sample when one accounted for the variation in the signal
to noise. Recently, a modest change with redshift has been observed
\citep{sullivan09}. These authors speculate that this variation is
simply due to the enhanced star formation rate at high redshifts and
thus one is looking at a sample with younger progenitors.

\begin{figure}
\includegraphics[width=.65\textwidth,angle=0,clip]{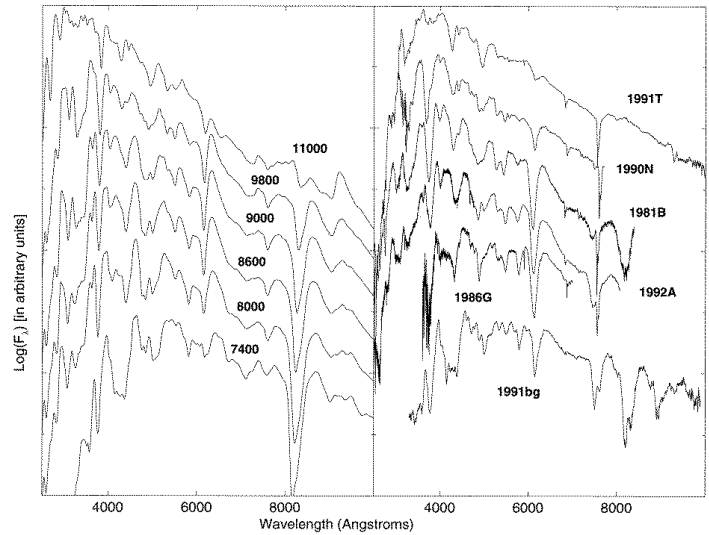}
\caption{The observed variation in spectra with brightness at maximum
  light is well reproduced by variation the model temperature in
  synthetic spectra calculations.}
\label{nugseq}
\end{figure}

\begin{figure}
\includegraphics[width=.65\textwidth,angle=0,clip]{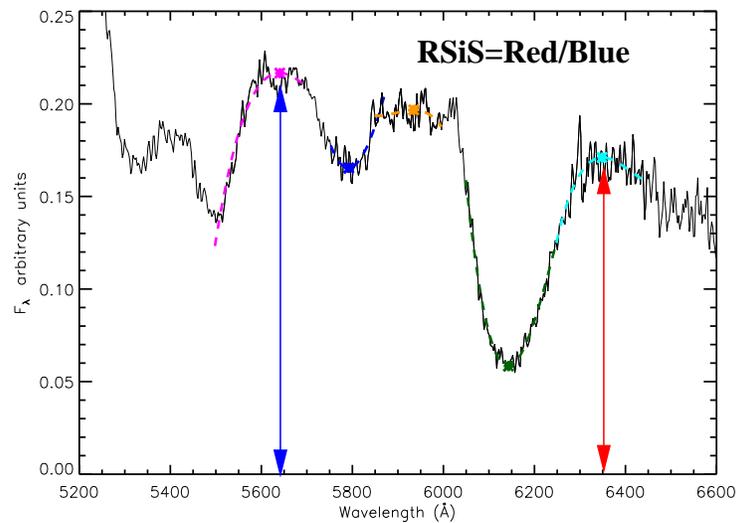}
\caption{The definition of the spectral ratio \RSiSu.}
\label{rsisdef}
\end{figure}

Regardless of whether or not the diversity can be corrected for using
purely empirical methods, an understanding of the diversity is needed.
To first order the diversity in the peak brightness has been
understood due to a variation in the total amount of nickel that is
produced in the explosion \citep{kmh93}. This is understood physically
as increasing the temperature which then leads to the observed
variations in the spectra \citep{nugseq95}. Figure~\ref{nugseq} shows
that simply varying the model temperature does a good job of
reproducing the observed variation. The spectral sequence also led to
the discovery that there are spectral indicators that are also well
correlated with the brightness at maximum light \citep{nugseq95,garn99by04}. 
These spectral indicators can be used as complementary to light curve
shape methods and since they have relatively small wavelength
baselines they are rather insensitive to properties of dust in the
parent galaxy. Figure~\ref{rsisdef} shows the definition of the ratio
\RSiSu\ which \citet{bongard06a} showed  can be 
used by JDEM. Recently, the SNfactory \citep{bailey09} showed that a
spectral ratio can be defined that reduces the variation in the Hubble
diagram to 12\%.

\subsection{3-D Explosions }

The explosion of a SN Ia is inherently three dimensional since the
white dwarf becomes convective shortly before it ignites and the
ignition is likely to be off-center. The flame also quickly becomes
wrinkled and thus there will be non-spherical compositions. Overall, 
SNe Ia are basically round, but explosion asymmetries
    lead to compositional and ionization inhomogeneities.  
Core-collapse supernovae almost certainly come from asymmetric
    engine. How important the  asymmetry is to spectra formation
    depends on how much of the hydrogen or helium envelope has been
    lost and how late one observes the 
    spectrum.  Gamma ray bursts are  beamed and highly
    relativistic. And of course AGN are asymmetric and may require
    general relativity.

\section{Detailed Quantitative Spectroscopy}

\subsection{\phx}

\phx is a generalized model atmosphere code, that solves the
generalized stellar atmosphere problem in static or moving flows using
the fully special relativistic approach. For supernovae with
characteristic velocities of 10,000~\kmps and maximum velocities of up
to 60,000~\kmps the exact special relativistic formulation is
preferred. \phx is well calibrated on many astrophysical objects
including the sun \citep{short05}, cool stars \citep{helling08}, hot
stars \citep{aufden06}, planets \citep{hbb08}, novae \citep{petz05},
and all types of supernovae \citep{bongard08,ketchum08,bprd07,bbh07,knop0703z}.

\begin{figure}
\includegraphics[width=.6\textwidth,angle=0]{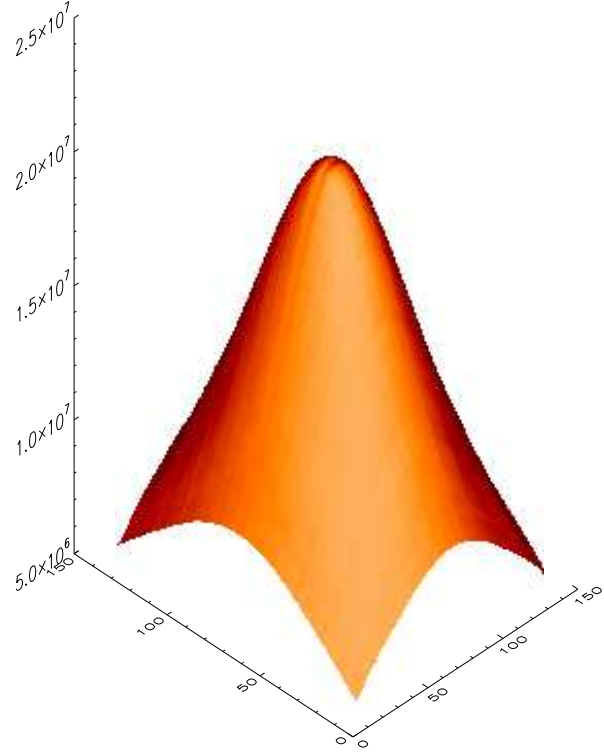}
\caption{Sphere in a box, mean intensity $J$ is plotted for a 
   a slice through the x-y plane for the case of a scattering
   continuum with $\epsilon = 10^{-4}$.}
\label{fig:sphereinbox}
\end{figure}

Over the last four years or so we have been developing a fully 3-D
version of \phx \citep{hb06,bh07,hb08,hb09a,bhb09,bin07}. We have
built this up slowly and carefully making sure to do careful code
verification along the way. Fig.~\ref{fig:sphereinbox} shows the
results from a spherically symmetric static test case with
scattering. This test was especially useful because it showed that our
full characteristics method was far superior to a short
characteristics method in terms of reproducing the spherical
symmetry. Since this model is a sphere in a box, the sphere is
surrounded by vacuum and the interpolations needed for the short
characteristics method did a very poor job of reproducing the correct
results across the opacity discontinuity.

\begin{figure}
\includegraphics[width=0.80\hsize,angle=0]{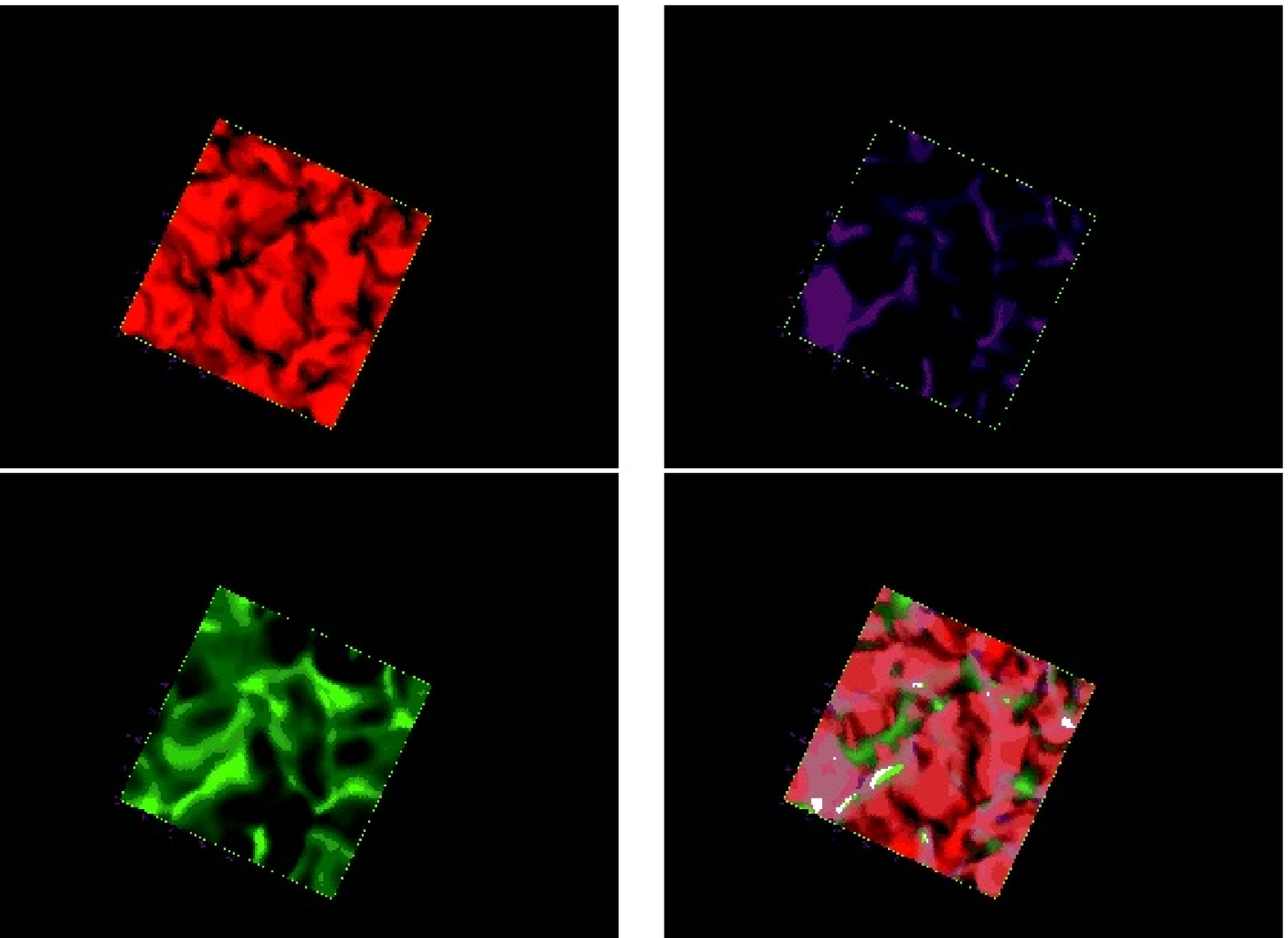}
\caption{The top left panel shows the continuum,
  the top right panel the line center,  the bottom left the line wing,
  and the bottom 
  right a composite image.}
\label{fig:solarmodel}
\end{figure}

Figure~\ref{fig:solarmodel} shows  a toy solar model  using periodic
boundary conditions. 
The temperature and density structure was taken from the model of
\citet{caffau07} but the opacity was taken to be proportional to the
density and independent of temperature. The thermalization parameter
in this calculation was taken to be
$\epsilon = 1$. 

Our general method is described in detail in
Refs~\citep{hb06,bh07,hb08,hb09a} and is modeled on the methods in our
1-D code \citep{phhs392,hbjcam99,hbmathgesel04}. 
The mean intensity $J$ is obtained from the source function
$S$ by a formal solution of the RTE which is symbolically written
using the $\Lambda$-operator $\Lambda$ as
\begin{equation}
     J = \Lambda S.              \label{frmsol}
\end{equation}
The source function is given by $S=(1-\e)J + \e B$, where $\e$ 
denotes the thermal coupling parameter and $B$ is Planck's function.

The $\Lambda$-iteration method, i.e.\ to solve Eq.~\ref{frmsol} by a fixed-point
iteration scheme of the form
\bea
   \Jnew = \Lambda \Sold , \quad
   \Snew = (1-\e)\Jnew + \e B  ,\label{alisol}
\eea
fails in the case of large optical depths and small $\e$.
The idea of the ALI or operator splitting (OS) method is to reduce the 
eigenvalues of the amplification matrix in the iteration scheme
\citep{cannon73}  by 
introducing an approximate $\Lambda$-operator (ALO) $\lstar$
and to split $\Lambda$ according to
\begin{equation}
           \Lambda = \lstar +(\Lambda-\lstar) \label{alodef}
\end{equation}
and rewrite Eq.~\ref{alisol} as
\begin{equation}
     \Jnew = \lstar \Snew + (\Lambda-\lstar)\Sold. 
\end{equation}
This relation can be written as \cite{hamann87}
\begin{equation}
    \left[1-\lstar(1-\e)\right]\Jnew = \Jfs - \lstar(1-\e)\Jold, \label{alo}
\end{equation}
where $\Jfs=\Lambda\Sold$  and $\Jold$ is the current
estimate of the mean intensity $J$. Equation~\ref{alo} is solved to
get the new values of  
$\Jb$ which is then used to compute the new 
source function for the next iteration cycle.

\begin{figure}
\includegraphics[height=0.35\textheight,angle=90]{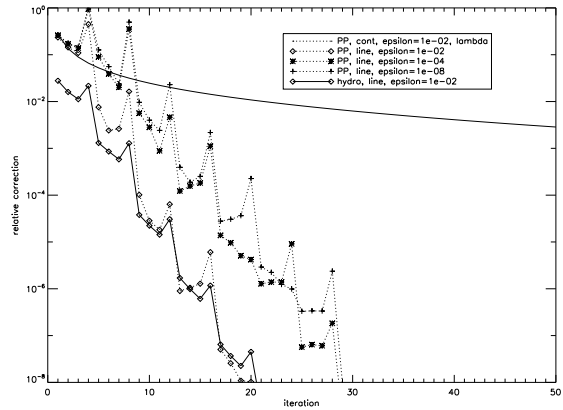}
\caption{Convergence rates of the 3D transfer for
line transfer with plane-parallel test structures (label `PP') and the 3D
hydro structure (label `hydro'). For comparison, the convergence of
the $\Lambda$ iteration for plane-parallel continuum transfer is 
also shown.}
\label{fig:convergence}
\end{figure}

Matrix computations are required in order to solve Eq.~\ref{alo}. In
choosing matrix methods one must remember that the  inverse of a
banded matrix is full. 
Band matrix solvers require large amounts of memory. This is also the
case 
  for parallelized band matrix solvers.
 Iterative methods (Jordan and Gauss-Seidel) work well and use
  little memory. Ng acceleration is still very
  useful. Figure~\ref{fig:convergence} shows that the ALO works well
  and the Ng acceleration is evident.

\section{Moving Atmospheres}

For the moving case one must derive the equation of radiative transfer
using the techniques of general relativity. The pioneering work in
this field  was  done by
Lindquist \citep{lind66}. We sketch the development of the solution of
the relativistic radiative transfer equation beginning with the work
of Mihalas \citep{mih80} and describe the work we have done
\citep{bhb09,bin07}. 

\subsection{Mihalas' Method}

In his important paper Dimitri Mihalas \citep{mih80} explained why we
want to work in the co-moving frame (italics as in original):
\begin{quote}
The emissivity $\eta$ and opacity $\chi$ depend upon the angle as
well as frequency in the inertial frame because of Doppler shifts,
aberration, and advection induced by the motion of the material in the
frame.

The goal of this section is to rewrite equation (2.1) with \emph{all
  material and radiation-field quantities measured in the comoving
  frame}; in that frame both the opacity and emissivity are isotropic,
and can be related directly to proper variables that specify the
thermodynamic state of the material. Furthermore, in that frame both
the scattering properties of the material and the rate equations
describing the mechanisms populating and depopulating its internal
energy states are most easily defined\dots

In our analysis we shall, however, leave both the \emph{space and time
  variables in the inertial frame}, as this is the only frame in which
synchronism of clocks can be effected, and further this choice
obviates the need to develop a metric for accelerated fluid frames
\citep{castor72} which in general can only be done approximately. With
this choice of frame we can write exact Lorentz transformations for
all the material and radiation-field quantities and use these to
develop a transfer equation that will remain valid for relativistic
flow in the limit as $v/c \longrightarrow 1$\dots
\end{quote}

In \citet{mih80} the characteristic equations are written as coupled
ordinary differential equations in both  the {\bf inertial} frame
spatial coordinate, $r,$ and the 
{\bf co-moving} frame momentum space coordinate, $\mu$, i.e.,
\begin{eqnarray*}
\frac{dr}{ds_M} &=& \gamma(\mu+\beta), \\ 
\frac{d\mu}{ds_M}&=&\gamma(1-\mu^2)\left[{1+\mu\frac{\beta}{r}}-
\gamma^2(\mu+\beta)\frac{d\beta}{dr}\right].
\end{eqnarray*} 
From these results of Mihalas a generation of
workers in the field have been educated that the
characteristics are curved in phase space \citep[see Figure 1 of
Reference][]{mih80}.  

\subsection{The Affine Method}

\begin{figure}
  \includegraphics[height=.3\textheight]{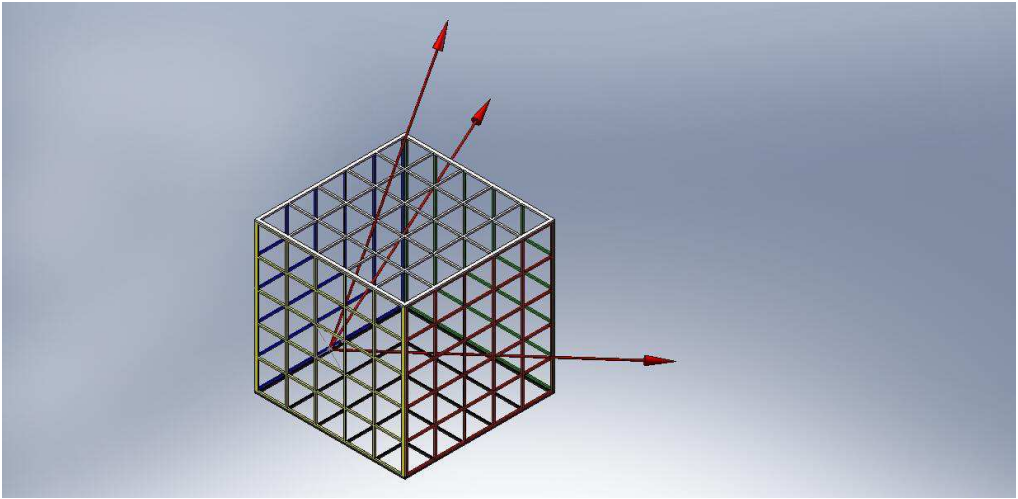}
  \caption{Characteristics are straight lines in flat spacetime and
    they begin at a boundary point and traverse the computational
    volume with a constant direction measured in the observer's frame.}
\label{fig:charrays}
\end{figure}

Generalizing the method of \citet{mih80} to 3 spatial
dimensions (and thus 6 phase space dimensions) is very
difficult. Following the methodology of \citet{mih80} 
we would keep all 3 momentum space variables \textbf{co-moving}. Then, in
order to keep track of the evolution of those co-moving phase space
variables along the characteristics (e.g., $\mu,$ $\phi$), we would
have relied on the so-called tetrad formalism
\citep{morita84,morita86,castor09} which is very complicated for
spacetimes with no symmetry or arbitrary flows. For an example, a boost
followed by an arbitrary $SO(3)$ rotation is still a valid Lorentz
transformation and thus it is difficult to define a consistent frame
in the case of arbitrary flows.  In \citet{bin07} we realized that photons
travel along null geodesics (straight lines in flat spacetime), which
depends only on the background spacetime, not the velocity
distribution of the flow, and consequently is either known
analytically or can be numerically solved before we solve the RTE. The
so-called characteristics in phase space are just the `unique'
lifting of the geodesics in the $4$-D spacetime manifold to the $8$-D
tangent bundle of the spacetime manifold \citep{lind66,ehlers71a}. The
apparent curvature of the characteristics in \citep{mih80} is due to
the selection of phase space coordinates. To avoid the complications
induced by the use
of the tetrad formalism, we found that we could work in a very special
co-moving frame where only the wavelength of the photon is measured by
a co-moving observer (which differs from the inertial frame value by
only a Doppler factor and is thus easily calculated), the other two
momentum directions are measured in the observer's frame (which in
flat spacetime are constants). This is not a giant leap, since in the
case of \citet{mih80} the transfer equation is solved in a mixed
frame, where the spatial variables are Eulerian, but the two momentum
variables $\mu$ and $\lambda$ are measured by a co-moving observer.
Figure~\ref{fig:charrays} illustrates the characteristics for a
Cartesian grid.

The above considerations lead to a radiative transfer equation given by
\be
\left.\frac{\partial I_\lambda}{\partial s}\right|_\lambda + a(s)\lambda\frac{\partial I_\lambda}{\partial \lambda}=-[\chi_\lambda f(s)+5 a(s)]I_\lambda+\eta_\lambda f(s).
\label{rte}
\ee
where the notation is the same as in \citet{bhb09}.

The evaluation of $I_\lambda$ in this mixed frame makes it quite easy
to solve the scattering problem in the co-moving frame. To get the
co-moving $J_\lambda,$ we need to integrate the (co-moving) specific
intensity $I_{\lambda}(\bf{r},\hat{\bf n})$ over the co-moving solid
angle element $d\Omega.$ Since $I_{\lambda}(\bf{r},\hat{\bf n})$ is
given in terms of the {\bf inertial} frame direction $\hat{\bf n},$ it
is desirable to transform the integral into one over the {\bf
  inertial} frame solid angle element $d\Omega_0,$ this causes no
difficulty since
\begin{eqnarray}
d\Omega &=& (\gamma[1 - \mathbf{\beta\cdot  n}])^{-2} d\Omega_0
\nonumber \\
 &=& f(s)^{-2}d\Omega_0. \label{tocomov}
\end{eqnarray}
Using Eq.~\ref{tocomov}, $J_\lambda$ in the co-moving frame can be
expressed as 
\begin{equation}
J_\lambda = \int I_\lambda({\bf r},{\hat{\bf n}}) f(s)^{-2} d\Omega_0
\end{equation}
where $I_\lambda({\bf r},{\hat{\bf n}})$ is expressed in the ``funny
frame'' that comes from solving Eq.~\ref{rte}.

\begin{figure}
   \includegraphics[width=0.65\hsize]{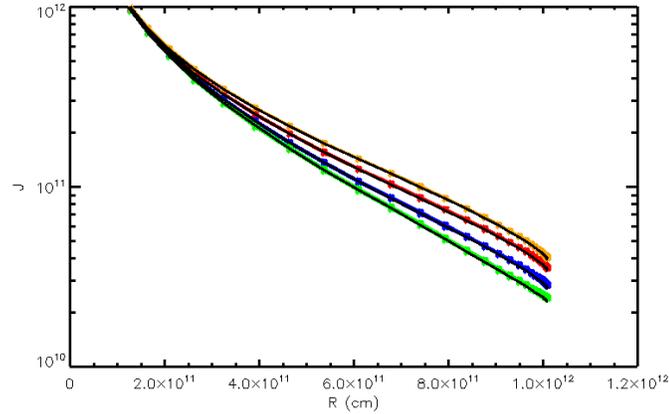}
    \caption{The results of 1-D calculations are compared with 3-D
   calculations for $\beta_{max} = (0.03,0.33,0.67,0.87)$.}
\label{fig:highv}
\end{figure}

\begin{figure}
   \includegraphics[width=0.65\hsize]{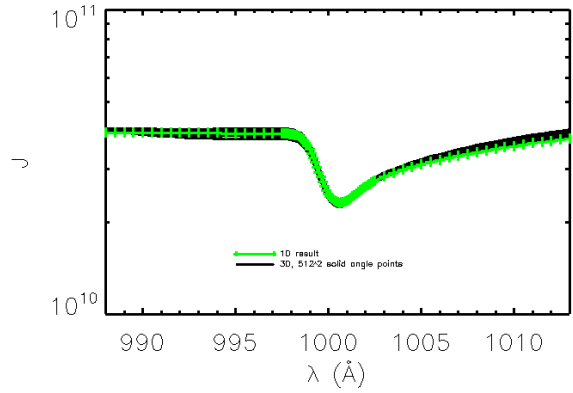}
\caption{The mean intensity of a line in the co-moving frame for a line
  with $\epsilon = 0.1$, $\beta_{max} = 0.03$. The solid lines are the
3-D result and the plus signs are the 1-D results.}
\label{fig:line_comov}
\end{figure}

\begin{figure}
   \includegraphics[width=0.65\hsize]{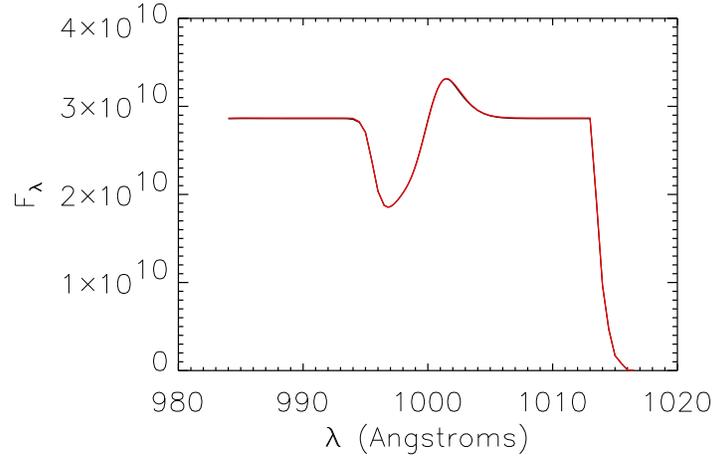}
\caption{The line shown in Fig~\protect{\ref{fig:line_comov}}
  transformed to the observer's frame gives the expected P-Cygni
  profile.}
\label{fig:line_obs}
 \end{figure}

\begin{figure}
   \includegraphics[width=0.65\hsize,angle=90]{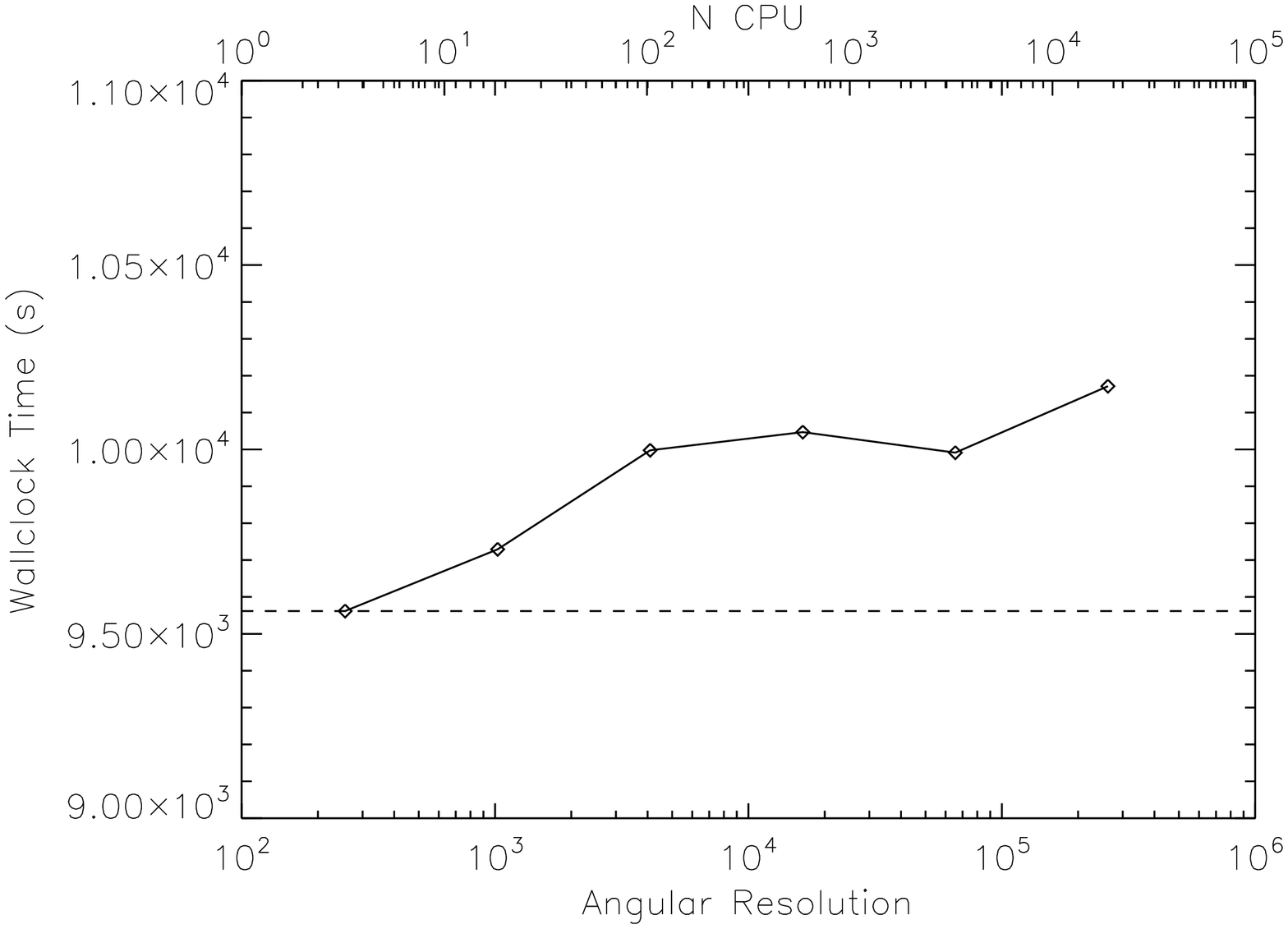}
    \caption{Wall-clock time for $\epsilon
        = 0.1$  as a function of angular resolution/number of CPUs.}
      \label{fig:speedup}
    \end{figure}

We have implemented this method in spherical coordinates for the case
of homologous flows \citep{bhb09}. It is important to verify that the
code gives the correct results in a case that can be tested. Luckily,
we can compare the results for a spherically symmetric test with those
of our well-tested 1-D code, which uses the Mihalas' method and not
the affine method. Figure~\ref{fig:highv} compares the results for the
mean intensity $J$ in the co-moving frame of the
1-D code (dots) and the 3-D  (solid lines) at each voxel at a given
radius for 
differing maximum velocities. No scaling is performed between the two
calculations. The agreement is excellent and the 3-D code does a very
good job of reproducing spherical symmetry. This also shows that there
is no problem in the interpretation of our
frame. Figure~\ref{fig:line_comov} shows similar results for a
scattering line in the co-moving frame.  Figure~\ref{fig:line_obs}
shows the flux in the observer's frame reproduces the expected P-Cygni profile.
Figure~\ref{fig:speedup} shows that the code parallelizes
quite well in terms of resolution. The scaling test has each processor
working on the same number of characteristics as the number of
momentum space angles is increased. The 14\% cost is reasonable given
the increased communication overhead as one goes from 256 to 16,384
processors.

\section{Summary}

  \phx now includes tested 3-D fully relativistic radiative transfer
  with homologous flows.  
  Including arbitrary flows is a computational, not an algorithmic
  challenge \citep{bh04,khb09} and we are working on this.
  The next step is to go beyond test problems to a full production code.
  We expect progress from the inter-comparison of new datasets of nearby
  supernovae 
    with 3-D hydro models and synthetic spectroscopy. The affine
    method is extremely easy to implement and can be applied to
    radiation hydro codes. Happy birthday Dimitri!

\begin{theacknowledgments}
This work was supported in part NSF grant AST-0707704,  US DOE Grant
DE-FG02-07ER41517 and   SFB 676
from the DFG.    This research used resources of the National Energy
Research Scientific Computing Center (NERSC), which is supported by the Office
of Science of the U.S.  Department of Energy under Contract No.
DE-AC02-05CH11231; and the H\"ochstleistungs Rechenzentrum Nord (HLRN).  We
thank both these institutions for a generous allocation of computer time.
\end{theacknowledgments}

\bibliographystyle{aipproc}   
\bibliography{apj-jour,refs,baron,sn1bc,sn1a,sn87a,snii,stars,rte,cosmology,gals,agn,crossrefs}

\IfFileExists{\jobname.bbl}{}
 {\typeout{}
  \typeout{******************************************}
  \typeout{** Please run "bibtex \jobname" to optain}
  \typeout{** the bibliography and then re-run LaTeX}
  \typeout{** twice to fix the references!}
  \typeout{******************************************}
  \typeout{}
 }

\end{document}